\documentclass[12pt]{iopart}
\usepackage{times}
\usepackage{graphicx}
\begin{document}

\title{Combinatorics of Boson Normal Ordering: the Dobi\'{n}ski Formula Revisited}
\author{Karol A. Penson\footnote{penson@lptl.jussieu.fr} and Allan I. Solomon\footnote{a.i.solomon@open.ac.uk}}
\address{Laboratoire de Physique Théorique des Liquides,\\Université Paris VI, 75252 Paris Cedex 05,
  France}
\begin{abstract}
We derive explicit formulas for the normal ordering of powers of arbitrary monomials of boson operators. These formulas lead to generalisations of conventional Bell and Stirling numbers and to appropriate generalisations of the Dobi\'{n}ski relations.  These new combinatorial numbers are shown to be coherent state matrix elements of powers of the monomials in question. It is further demonstrated that such Bell-type numbers, when considered as power moments, give rise to positive measures on the positive half-axis, which in many cases can be written in terms of known functions. 
\end{abstract}

The standard boson commutation relation 
$
[a,a^{\dagger}]=1
\label{bos}
$
can be realised by identifying {\em formally} $a= \frac{d}{dx}$ and $a^{\dagger}=x$, since 
 $
[\frac{d}{dx},x] = 1.
\label{d}
$
In the present note we shall use both forms.  Integer sequences arise naturally  when considering the action of 
$(x \, {d}/{dx})^{n}$ on $f(x)$ 
which in general can be written as 
\begin{equation} 
(x\,\frac{d}{dx})^{n}f(x) =\sum_{k=1}^{n} S(n,k) x^{k} ({d}/{dx})^{k}f(x)
\label{s1}
\end{equation}
or, alternatively \cite{kat1}
\begin{equation} 
(a^{\dagger}a)^{n} =\sum_{k=1}^{n} S(n,k) (a^{\dagger})^{k} a^{k}.
\label{s2}
\end{equation}
The {\em Stirling numbers of the second kind} $S(n,k)$ appearing in Eqs.(\ref{s1}) and (\ref{s2}) have been known for over 250 years \cite{yab}.  Eq.(\ref{s2}) exemplifies the {\em normal ordering problem}, that is, finding the form of $(a^{\dagger}a)^{n}$ with the powers of $a$ on the right. Although explicit expressions for $S(n,k)$ are known \cite{com}, of particular interest here are the {\em Bell numbers} $B(n)$ given by the sums
\begin{equation} 
B(n) =\sum_{k=1}^{n} S(n,k), \; \; \; \; \; \; \; n=1,2,\ldots
\label{b1}
\end{equation}
with $B(0)=1$ by convention $(S(n,0)=\delta_{n,0})$. A closed-form expression for $B(n)$ can be found by considering the action of $(x {d}/{dx})^{n}$ on $f(x)=e^{x}$ giving:
\begin{equation} 
(1/e^{x}) \sum_{k=0}^{\infty} {\frac{k^n}{k!}}x^{k} =\sum_{k=1}^{n} S(n,k)x^{k},
\label{b2}
\end{equation}
which for $x=1$ reduces to
\begin{equation} 
(1/e) \sum_{k=0}^{\infty} \frac{k^n}{k!} =\sum_{k=1}^{n} S(n,k)=B(n).
\label{b3}
\end{equation}
Equations (\ref{b2}) and (\ref{b3}) are the celebrated Dobi\'{n}ski formulas (\cite{yab},\cite{com},\cite{con},\cite{wil}) which have been the subject of much combinatorial interest.  Eq.(\ref{b3}) represents the integer $B(n)$ as an infinite series, which is however {\em not} a power series in $n$.  An immediate consequence of Eqs.(\ref{b2}) and (\ref{b3}) is that $B(n)$ is the $n$-th moment of a (singular) probability distribution, consisting of weighted Dirac delta functions located at the positive integers (the so-called Dirac comb) :
\begin{equation} 
B(n) = \int_{0}^{\infty} x^{n} W(x) dx, \; \; \; n=0,1,\ldots
\label{d1}
\end{equation}
where
\begin{equation} 
W(x) =(1/e) \sum_{k=1}^{\infty} {\frac{\delta(x-k)}{k!}}.
\label{d2}
\end{equation}
The discrete measure $W(x)$ serves as a weight function for a family of orthogonal polynomials ${\cal C}_n^{(1)}(x)$, the Charlier polynomials \cite{koe}. The exponential generating function (EGF) of the sequence $B(n)$ can be obtained from Eq.(\ref{b3}) as
\begin{equation} 
e^{(e^{\lambda} -1)} =\sum_{k=0}^{\infty} {B(n) \frac{{\lambda}^n}{n!}}.
\label{EGF}
\end{equation}
This equation  is related via Eq.(\ref{s2}) to a formula giving  the normal ordered form of $e^{\lambda a^{\dagger} a}$ \cite{kla},\cite{lou}
\begin{equation}
e^{\lambda a^{\dagger} a} = {\cal{N}}(e^{\lambda a^{\dagger} a}) = :e^{a^{\dagger} a(e^{\lambda}-1)}: 
\label{norm}
\end{equation} 
 The symbol ${\cal{N}}$ denotes normal ordering, while $: :$ refers to normal-ordering with neglect of non-commutativity. We stress that in the derivation 
of  Eq.(\ref{norm})  in \cite{kla} and \cite{lou} no use has been made of the Stirling and Bell numbers. It may readily be seen that Eq.(\ref{EGF}) is the expectation value of Eq.(\ref{norm}) in the coherent state $|z\rangle$ defined by $a|z\rangle=z|z\rangle$ at the value $|z|=1$, using Eqs.(\ref{s2}) and (\ref{b1}).
This circumstance has been used recently to re-establish the link between the matrix element $\langle z|e^{\lambda a^{\dagger} a}|z\rangle$ and the properties of Stirling and Bell numbers \cite{kat}.

The purpose of this note is to show that the above results on functions of $a^{\dagger}a$ may be extended to functions of $(a^{\dagger})^{r} a^{s}, \;  (r,s = 1,2,\ldots) $ with $r\geq s$; thus extending the  Dobi\'{n}ski relations. 

With this in mind we pose the following questions:
\begin{enumerate}
\item What extensions of the conventional Stirling and Bell numbers  occur in the normal ordering of $[(a^{\dagger})^{r} a^{s}]^{n}\;$?
\item Can the generalised Bell numbers $B_{r,s}(n)$ so defined be represented by an infinite series of the type of Eq.(\ref{d1}) - that is, do they satisfy a generalised Dobi\'{n}ski formula ?
\item May one consider the  $B_{r,s}(n)$ as the $n$-th moments of a positive weight function $W_{r,s}(x)$ on the positive half-axis, and may this latter be explicitly obtained ?
\end{enumerate}
In this note we indicate affirmative answers to these questions.

To this end we generalize  Eq.(\ref{s2}) by defining for $r \geq s$:
\begin{equation}
[(a^{\dagger})^{r} a^{s}]^{n}=(a^{\dagger})^{n(r-s)}\sum_{k=s}^{ns} S_{r,s}(n,k)(a^{\dagger})^{k}{a}^{k}  
\label{s3}
\end{equation}
or, alternatively,
\begin{equation}
[{x}^{r} (d/dx)^{s}]^{n}={x}^{n(r-s)}\sum_{k=s}^{ns} S_{r,s}(n,k){x}^{k}\,(d/dx)^{k}.
\label{s4}
\end{equation}
Eqs.(\ref{s3}) and ({\ref{s4}) introduce  generalized Stirling numbers $S_{r,s}(n,k)$ which imply an extended definition of generalized Bell numbers:
\begin{equation}
 B_{r,s}(n) \equiv \sum_{k=s}^{ns} S_{r,s}(n,k).
\label{eb}
\end{equation}
Note that the Stirling numbers $S_{r,1}(n,k)$ were studied in \cite{lan}. Also $B_{1,1}(n)=B(n)$ of Eq.(\ref{b1}). 

We have found a representation of the numbers $B_{r,s}(n)$ as an infinite series, which is a generalization of the Dobi\'{n}ski formula Eq.(\ref{b3}).  For $r=s$ one obtains:
\begin{equation}
B_{r,r}(n) = (1/e) \sum_{k=0}^{\infty}\frac{1}{k!}\left[\frac{(k+r)!}{k!}\right]^{n-1}, \; \; \; n=1,2 \ldots
\label{b4}
\end{equation}
with $B_{r,r}(0) = 1$ by convention.
For $r>s$ the corresponding formula is:
\begin{equation}
B_{r,s}(n) = [{(r-s)^{s(n-1)}}/e ] \sum_{k=0}^{\infty} \left[\prod_{j=1}^{s} \frac{\Gamma(n+\frac{k+j}{r-s})}{\Gamma(1+\frac{k+j}{r-s})}\right], \; \; B_{r,s}(0)=1.
\label{b5}
\end{equation}

The formula Eq.(\ref{b3}) and its extensions Eqs.(\ref{b4}) and (\ref{b5}) share a common feature, namely, the fact that they give rise to a series of integers is by no means evident!

A general family of sequences arising from Eq.(\ref{b5}) has the form $(p,r=1,2,\ldots)$:
\begin{eqnarray}
\hspace{-2.5cm} B_{pr+p,pr}(n) & = & (1/e)\left[\prod_{j=1}^{r} \frac{(p(n-1)+j)!}{(pj)!}\right] \times \nonumber \\     
\hspace{-2.5cm}           &   & \times \,_{r}\!F_{r}(pn+1,\dots,pn+1+p(r-1);1+p,\ldots,1+p+p(r-1);1),
\label{b6}
\end{eqnarray}
where $_{r}\!F_{r}$ is the hypergeometric function.
Knowledge of the generalized Stirling numbers in Eq.(\ref{s3})  solves the normal ordering problem for $[(a^{\dagger})^{r} a^{s}]^{n}\,$. We are able to give the appropriate generating functions for the sequences $B_{r,s}(n)$.  It then follows that, at least formally, we can furnish the generating functions for $S_{r,s}(n,k)$ as well \cite{pen}.  Additionally, it turns out that in certain circumstances  one may obtain explicit expressions for them.  We quote two such cases:
\begin{equation}
S_{r,r}(n,k)=\sum_{p=0}^{k-r} \frac{(-1)^{p}[\frac{(k-p)!}{(k-p-r)!}]^{n}}{(k-p)!p!} \;\;\;\; (r\leq k \leq rn)
\end{equation}
and
\begin{equation}
S_{2,1}(n,k)=\frac{n!}{k!}\left(\matrix{n-1 \cr k-1\cr}\right) \; \; \;  (1 \leq k\leq n)
\end{equation}
which are the so-called unsigned Lah numbers \cite{com,lan}.
  
For those pairs $(r,s)$ for which we have an explicit expression for $S_{r,s}(n)$ we may generalize  Eq.(\ref{norm}) to obtain the normal ordered form of $e^{\lambda (a^{\dagger})^{r} a^{s}}$.  For example, the matrix element 
$\langle z |e^{\lambda (a^{\dagger})^{r} a}|z \rangle$   leads to the following normally ordered expression:
\begin{equation}
e^{\lambda (a^{\dagger})^{r} a} = {\cal{N}}(e^{\lambda (a^{\dagger})^{r} a}) = :\exp\{ [ (1-\lambda (a^{\dagger})^{r-1} (r-1))^{\frac{1}{r-1}} -1] a^{\dagger} a\}: 
\label{norm2}
\end{equation}
We apply the method  to Eq.(\ref{norm2}) which gave the EGF of Eq.(\ref{EGF})  and take the expectation of Eq.(\ref{norm}) in the coherent state $|z\rangle$ to get:
\begin{equation}
\langle z |e^{\lambda (a^{\dagger})^{r} a} |z\rangle = :\exp\{ [ (1-\lambda (z^{*})^{r-1} (r-1))^{\frac{1}{r-1}} -1] |z|^{2}\}: 
\label{expec}
\end{equation}
which evaluates at $z=1$ giving
\begin{equation}
\langle z |e^{\lambda (a^{\dagger})^{r} a} |z\rangle{_{z=1}} = :\exp\{  (1-\lambda  (r-1))^{\frac{1}{r-1}} -1 \}: \label{EGF2}
\end{equation}
which is precisely the EGF for the numbers $B_{r,1}(n)$ \cite{lan}.

For general $r>s$ the corresponding  $B_{r,s}(n)$ grow much more rapidly than $n!$ and thus may not be obtained via the usual form of EGF. One instead defines the EGF in terms of  $B_{r,s}(n)/(n!)^{t}$ where $t$ is an integer chosen to ensure that $\sum_{n=0}^{\infty}{B_{r,s}(n) }/{(n!)^{t+1}}$ has a finite radius of convergence.  As a result one obtains variants of Eq.(\ref{expec}) involving different hypergeometric functions \cite{pen}.

The analogue of Eq.(\ref{d1}) in the general case 
\begin{equation}
 B_{r,s}(n)=\int_{0}^{\infty} x^{n} W_{r,s}(x) dx 
\label{sti}
\end{equation}
leads to  weight functions $W_{r,s}(x)$ which may be shown to be positive through use of properties of the Mellin transform, and for some of which analytic expressions may be obtained.  For $r=s$ we obtain discrete measures, giving a "rarefied" form of the Dirac comb, while for $r>s$ we obtain continuous measures.
We conclude this note with one  example of the latter kind:
\begin{equation}
W_{2r,r}(x)= \frac{1}{e \, r}x^{\frac{2-3r}{2r}}e^{-x^{\frac{1}{r}}}I_{r}(2x^{\frac{1}{2r}})
\end{equation}
where $I_{1}(y)$ is the modified Bessel function of the first kind.  
% This is how we recommend you do the references in LaTeX


\section*{References}
\begin{thebibliography}{99}
% The next line is there to match the Word file
% By default the references are printed smaller in the iopart style
\normalsize
\bibitem{kat1}
Katriel J and Duchamp G 1995
J. Phys. A {\bf 28} 7209
\bibitem{yab}
Yablonsky S V 1989
Introduction to Discrete Mathematics
(Moscow: Mir Publishers) 
\bibitem{com}
Comtet L 1974
Advanced Combinatorics
(Dordrecht: Reidel) 
\bibitem{con}
Constantine G M and Savits T H 1994
SIAM J. Discrete Math. {\bf 7} 194.
\bibitem{wil}
Wilf H 1994
Generatingfunctionology
(New York: Academic)
\bibitem{koe}
Koekoek R and Swarttouv R F 1998
The Askey scheme of hypergeometric polynomials and its q-analogue
Dept. of Technical Mathematics and Informatics, Report No. 98-17
Delft University of Technology
\bibitem{kla}
Klauder J R and Sudarshan E C G 1968
Fundamentals of Quantum Optics
(New York: Benjamin)
\bibitem{lou}
Louisell W H 1977
Radiation and Noise in Quantum Electronics
(Florida: Krieger)
\bibitem{kat}
Katriel J 2000
Phys. Lett. {\bf A273} 159
\bibitem{lan} Lang W 2000
J. Int. Seqs. {\bf 12} Article 00.2.4
(www.research.att.com/\~{}njas/sequences/)
\bibitem{pen}
Penson K A and Solomon A I 2002
to be published.
\end{thebibliography}
\end{document}